\documentclass[useAMS,usenatbib]{mn2e}

\usepackage{amssymb}
\usepackage{graphicx}
\usepackage{natbib}

\title[Radial stability of anisotropic Hernquist models]
{Radial stability of a family of anisotropic Hernquist models with and without a supermassive black hole} 

\author[Buyle, Van Hese, De Rijcke \& Dejonghe]{%
  P. Buyle$^{1}$\thanks{E-mail: Pieter.Buyle@UGent.be}, E. Van
  Hese$^1$, S. De Rijcke$^1$, and H. Dejonghe$^1$\\
  $^{1}$Sterrenkundig Observatorium, Ghent University, Krijgslaan 281,
  S9, B-9000 Ghent, Belgium}
\begin{document}

\date{}

\pagerange{\pageref{firstpage}--\pageref{lastpage}} \pubyear{2006}

\maketitle

\label{firstpage}

\begin{abstract}
We present a method to investigate the radial stability of a spherical
   anisotropic system that hosts a central supermassive black hole
   (SBH). Such systems have never been tested before for stability,
   although high anisotropies have been considered in the dynamical
   models that were used to estimate the masses of the central
   putative supermassive black holes. A family of analytical
   anisotropic spherical Hernquist models with and without a black
   hole were investigated by means of $N$-body simulations. A clear
   trend emerges that the supermassive black hole has a significant
   effect on the overall stability of the system, i.e.\ an SBH with a
   mass of a few percent of the total mass of the galaxy can prevent
   or reduce the bar instabilities in anisotropic systems. Its mass
   not only determines the strength of the instability reduction, but
   also the time in which this occurs. These effects are most
   significant for models with strong radial
   anisotropies. Furthermore, our analysis shows that unstable systems
   with similar SBH but with different anisotropy radii evolve
   differently: highly radial systems become oblate, while more
   isotropic models tend to form into prolate structures. In addition
   to this study, we also present a Monte-Carlo algorithm to generate
   particles in spherical anisotropic systems.
\end{abstract}


\newcommand{\rev}{\!\!\!}
\section{Introduction}
Nowadays it is accepted that almost every galaxy hosts a central
supermassive black hole (SBH) at its core. Since the kinematical
discovery of the first SBH with the Hubble Space Telescope (HST),
extensive studies have been carried out by many groups that
investigate the demography of SBHs and the effect of the SBHs on their
environment. The most popular discoveries are the correlations between
the mass of the SBH ($M_{BH}$) and respectively the total blue
magnitude $L_B$ of the hot stellar component in which it resides
\citep{korrich}, the central velocity dispersion of the hot stellar
component \citep{fermer,gebhardt}, the central light concentration
$C(\alpha)$ or equivalent the S\'ersic index $n$ \citep{graham} and
the maximum rotational velocity of the galaxy
\citep{fer,baes4,piz05,buyle06}. These relations have been calibrated
with the known masses of the SBHs of the nearest galaxies, that mostly
have been derived by means of either stellar or gas kinematics.

Sophisticated axisymmetric 3-integral dynamical models that allow a
variation in mass-to-light ratio and anisotropy as a function of
radius have been obtained by fits to the line-of-sight velocity
distributions (LOSVDs) in the galaxies, which were derived primarily
from high-resolution spectra taken with the HST
\begin{equation}
LOSVD\,(x,y,v_z)=\frac{1}{\rho_p}\int\rev\int\rev\int F(\bmath{r},\bmath{v})\,dz\,dv_x\,dv_y,
\end{equation}
where $F(\bmath{r},\bmath{v})$ denotes the stellar distribution
function (DF) and $\rho_p$ stands for the projected mass density at
position $(x,y)$. The accuracy of the applied dynamical models to the
observed stellar kinematics is still improving steadily and is
reflected on the complexity of the DFs. Despite this positive progress
on the dynamical front, very few anisotropic dynamical models of a
galactic nucleus have been tested for dynamical stability
\citep{laura2}. One of the reasons for this is the complexity of the
distribution functions, which are mostly numerically derived. Hence,
to simulate these numerical DFs one normally approximates numerically
the solution of the Jeans equations to derive the velocity dispersion
profile and then uses Gaussians to provide local velocity
distributions. It is now known from recent simulations of galactic
systems that this method causes serious numerical artifacts
\citep{kazan}.

So far the only {\it theoretical} analytical systems that contain a
SBH are derived by Ciotti (1996) and Baes et al. (2004, 2005) where
the attention is drawn primarily to the Hernquist model since this is
the best-known approximation to the S\'ersic profiles that are
observed in bulges and elliptical galaxies, and by Stiavelli (1998)
where the distribution function of a stellar system around an SBH is
derived from statistical mechanic considerations. Ciotti (1996)
initially starts with a 2-component system containing the luminous and
dark matter and creates both isotropic and anisotropic (based on the
Osipkov-Merritt strategy) systems. The dark matter halo (also
represented by a Hernquist model) can be transformed into a central
SBH by setting the core radius to zero.

In this article we present for the first time the results of a
dynamical stability investigation of spherical systems containing an
SBH, as a function of the mass of the SBH and the anisotropy radius of
the system. We fill focus primarily on the so-called radial orbit
instability of radially anisotropic, spherically symmetric stellar
systems \citep{he73,ci96}. A beautiful mathematical and physical
explanation for this instability can be found in Palmer (1994) and
Merritt (1987) and references therein. As a bar grows, it saps angular
momentum from the stars as they precess through the bar. Stars on
low-angular momentum orbits are trapped into resonance and strengthen
the bar, making it possible to also trap stars with higher angular
momentum into resonance and so on. Eventually, the triaxial force
field of the bar becomes dominant and the initially rosette-shaped
orbits in a spherical potential are transformed into box
orbits. Radial orbits and box orbits bring stars close to the center
of the galaxy where they can be diverted from their orbits by the
spherically symmetric force field of a central massive black
hole. This in turn may weaken the bar over time, or even prevent it
from forming in the first place, clearly proving the relevance of a
study such as the one presented here.

In Section 2 we describe a Monte-Carlo algorithm that
we developed to generate the initial conditions for the models,
together with our $N$-body code and technique for
investigating the stability. We
present in Section 3 the results of a stability investigation of
a family of anisotropic Hernquist models without an SBH, with
different anisotropy behavior \citep{baes}. In Section 4 we describe
the Osipkov-Merritt Hernquist models with a central SBH, introduced by
Ciotti (1996). We investigate the stability of these systems in detail
in Section 5, comparing them with the according models without an SBH.
We perform this in a 2-parameter space as a function of the anisotropy
radius $r_a$ and the mass of the central SBH $\mu$.  In Section 6 we
present our final results and conclusions.

\section{Computational method}
\begin{figure}
\centering
\includegraphics[width=9cm,clip]{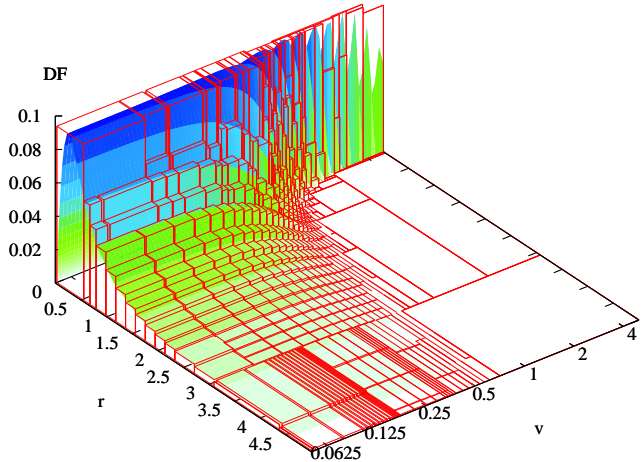}
    \caption{Visualization of an isotropic Hernquist system
with $\mu=0.1$ and our approximation with cells. After 8 subdivisions 991
cells were constructed, with a total phase-space volume of
$V_8=1.533$, while the real total stellar DF volume is 1.  Thus, the
ratio of rejected to accepted particles is $0.533$ and on
average $\sim 35\%$ of all randomly chosen test particles in the cell
volume will be rejected, resulting in a highly efficient Monte-Carlo
simulation.}
    \label{Figgrid}
\end{figure}
\subsection{Definition of the Hernquist models}
First of all we introduce some general characteristics of the models
in our dynamical study. All systems are based on the spherical
Hernquist potential-density pair \citep{hern1}, including a
supermassive black hole in the centre. Given this mass profile, we
shall investigate several distribution functions (DFs) consistent with
the density outside the center, and which we will refer to as the
stellar component.  If we denote the total stellar mass by $M_s$, we
can write the total mass as $M_\mathrm{tot} = M_s(1 + \mu)$, where the
fractional quantity $\mu$ determines the SBH mass $\mu M_s$. In our
subsequent analysis we will work in dimensionless units $G=M_s=1$, 
so that the gravitating binding potential and the density are given by
\begin{eqnarray}
\psi(r)&=&\frac{1}{1+r} + \frac{\mu}{r},\label{hernpsi}\\
\rho(r)&=&\frac{1}{2\pi}\frac{1}{r(1+r)^3} \quad(r > 0)\label{hernrho}.
\end{eqnarray}
We will also express the time-steps in our $N$-body code (the time
between two successive calculations) in dimensionless units of
half-mass dynamical time, which is defined as the dynamical time
\citep{binney87} at the stellar half-mass radius:
\begin{equation}
T_h = \sqrt{\frac{3\pi}{16G\bar{\rho}}}, 
\end{equation}
where
\begin{equation}
\bar{\rho} = \frac{3M(r_{1/2})}{4\pi r_{1/2}^3}.
\end{equation}
For a Hernquist model with $\mu=0$ the half-mass dynamical time and
the half-mass radius are
\begin{eqnarray}
T_h&=&\frac{\sqrt{2}}{2}\pi\left(1+\sqrt{2}\right)^{3/2},\\
r_{1/2}&=&1+\sqrt{2}.
\end{eqnarray}
We will also use these units for models with an SBH.

\begin{figure*}
\centering
\includegraphics[width=14cm,clip]{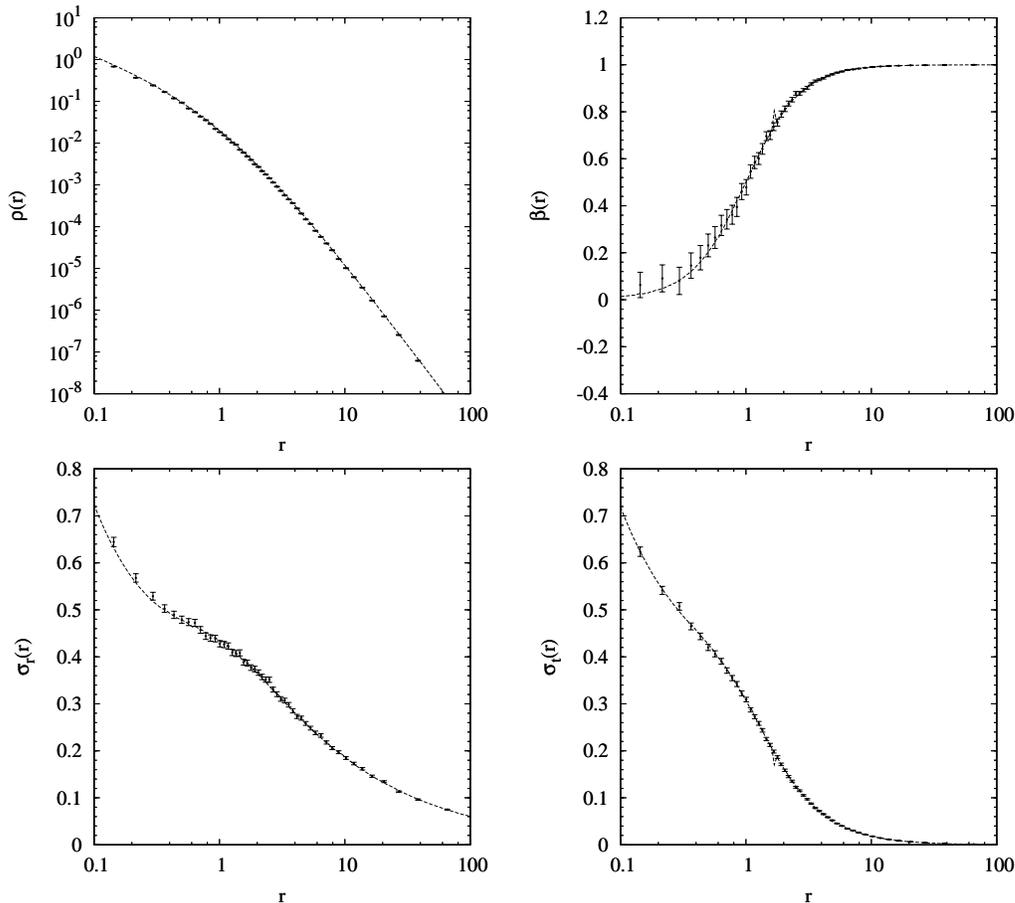}
    \caption{The figures show the relevant parameters ($\rho(r)$,
$\sigma_r(r)$, $\sigma_t(r)$ and $\beta(r)$) of the outcome of a Monte
Carlo sampling of a Hernquist system with $\mu=0.1$ and $r_a=1$. The
continuous lines denote the theoretical model, the discrete data
represent $10^5$ simulated particles, binned and with error-bars.}
    \label{Figsimulatie}
\end{figure*}

A conversion to observational units can be obtained through the close
similarity between the Hernquist and De Vaucouleurs profiles \citep{hern1},
with $r_{1/2} \approx 1.33 r_e$ where $r_e$ is the effective radius. Then a
physical length, time and velocity are found by the scaling relations
\begin{eqnarray}
\tilde{r}&=&\tilde{r}_u r,\\
\tilde{T}_h&=&\sqrt{\frac{\tilde{r}_u^3}{G \tilde{M}_s}}T_h,\\
\tilde{v}&=&\sqrt{\frac{G\tilde{M}_\mathrm{tot}}{\tilde{r}_u}}v,
\end{eqnarray}
with $\tilde{M}_s$ the stellar mass, $\tilde{M}_\mathrm{tot}$ the 
total mass and 
$\tilde{r}_u=(1.33r_e)/(1+\sqrt{2})$, expressed in physical units.

Every model was simulated by means of $10^5$ equal-mass particles that
all follow the distribution function of the system and are
contained within a sphere of radius $r_b=2000$ which encloses about
99.9\% of the stellar mass of the system. We performed the simulations
for 50 dynamical times, and used the  values of the axis ratios
$c/a$ and $b/a$ during this time as (in)stability indicators 
(see Section~\ref{axesstab}).

\subsection{Constructing the data sets}

Since we will investigate our models by means of $N$-body simulations,
the first objective is to obtain representative discrete data sets
from the considered distribution functions. Each of these functions
describes a spherical mass distribution in a dynamical system governed
by a gravitational binding potential $\psi(r) > 0$, which implies that
they can be expressed as functions $F(E,L)$ of the binding energy $E =
\psi(r) - \frac{1}{2}v_r^2 - \frac{1}{2}v_t^2$ and the modulus of the
angular momentum $L = rv_t$, with $v_r$ and $v_t$ the radial and
tangential velocity components, respectively. Isotropic DFs can be
reduced to $F(E)$.

In order to extract discrete data samples from the distributions, we
need to simulate random particles uniformly in the phase-space
enclosed by the DFs. To this aim we used a Monte-Carlo simulator,
developed by one of the authors (E.V.H.). The procedure works 
as follows: we write each DF as
$F(r,v_r,v_t)$  and we consider a 4-dimensional grid
space with $(r,v_r,v_t)$ as abscissae and the function
values on the ordinate axis.

We start with a single cell in this space, extending from the origin
to a boundary $(r_b,v_{r,b},v_{t,b})$ (where $r_b$ is chosen to be
sufficiently large, and $v_{r,b} = v_{t,b} = \sqrt{2\psi(0)}\,$), and
with the ordinate set at the (known or estimated) DF maximum
$f_b$. These boundaries (for infinite values a sufficiently large
value is taken, see further) enclose a 7-dimensional phase-space volume
\begin{equation}
V_1 = \bigg( \frac{4\pi}{3} r_b^3 \bigg) \bigg(2v_{r,b}^{}\bigg)\bigg(\pi v_{t,b}^2\bigg) f_b^{}.
\end{equation}

In the second step we attempt to split the cell into 8 sub-cells with
different ordinates (i.e.\ the up to that point known function maxima
in each cell). Therefore a co-ordinate $(r_s,v_{r,s},v_{t,s})$ is
sought to serve as the common corner point in the abscissae for these
sub-cells: starting in the cell center, the total phase-space volume of
the originating sub-cells is calculated, and through a number of
iterations the cell is scanned for a better splitting point, i.e.\
which minimizes this volume. In this manner, the original cell is
being split as efficiently as possible into 8 new cells, adding up to
a new total volume 
\begin{eqnarray}
V_2 \!\!\!\!&=&\!\!\!\! \sum_{i=1}^8 V_{2;i},\nonumber\\
 \!\!\!\!&=& \!\!\!\! \sum_{i=1}^8 \frac{8\pi^2}{3} (r_{b;i}^3 \!-\! r_{a;i}^3) 
(v_{r,b;i}^{} \!-\! v_{r,a;i}^{})(v_{t,b;i}^2 \!-\! v_{t,a;i}^2) f_{b;i}^{},
\end{eqnarray}
which is a better approximation to the real DF volume. Here, for a cell $i$ we
denoted $V_{2;i}$ its volume,
$(r_{a;i},v_{r,a;i},v_{t,a;i})$ and $(r_{b;i},v_{r,b;i},v_{t,b;i})$
its lower and upper bounds in the abscissae, and $f_{b;i}$ its maximum DF value.

Next, each cell in our grid is examined according to the procedure
above and split if it leads to a significant decrease in the total
volume. Thus, after the examination of every cell, a new volume $V_3$
is obtained.  This loop is repeated until after $M$ steps the
phase-space volume $V_M$ has converged sufficiently close to the real
volume. Typically, in our simulations, the cells cover a volume that
is a factor 1.5 to 5 larger than the model's actual phase-space
volume; a further refinement is unnecessary, since constructing more
cells would be more time-consuming than actually generating
our desired number ($10^5$) of particles (see below).
If the grid is successfully constructed, $F(r,v_r,v_t)$ is
entirely enveloped by a set of 4-dimensional grid cells.

\begin{figure*}
\centering
\includegraphics[width=19cm,clip]{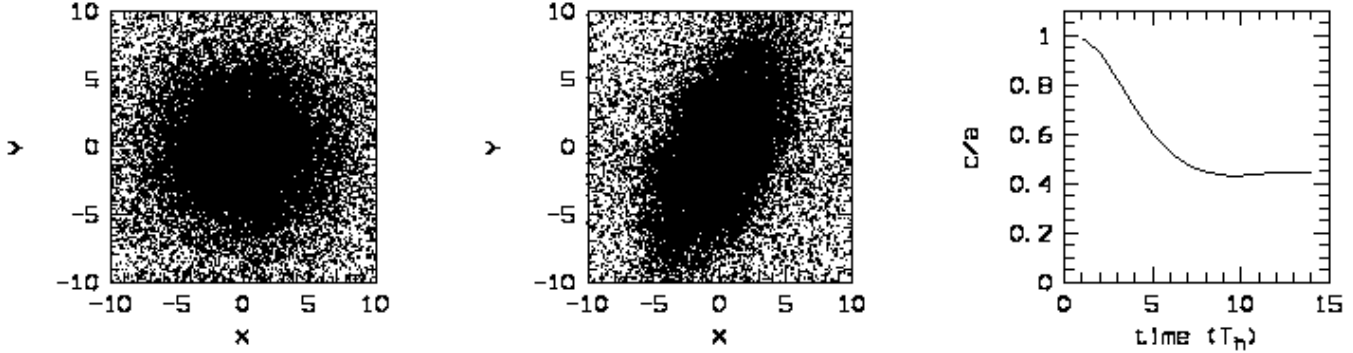}
    \caption{{\bf (a)} The initial particle positions of a Hernquist
    model with an increasing anisotropy with $\beta=0.5$ and
    $\lambda=5$. 80\% of the total mass is shown in the figure. {\bf
    (b)} The density distribution after 10 half-mass dynamical
    times. A bar is clearly visible. {\bf (c)} The axis ratio $c/a$
    plotted as a function of time. As can be seen from both the
    density distributions and the axis ratio, an elliptical bar is
    created indicating that the system is unstable.}
    \label{bar}
\end{figure*}

Now we can proceed to a classical rejection Monte-Carlo (MC)
simulation (in the remainder, we refer to setting up the initial
conditions of a DF as an ``MC simulation''). To generate a data
point $n$, first a value $V_n$ is randomly chosen between $0$ and $V_M$.
We can associate this value with a unique cell $j$ and an ordinate $f_n$
for which
\begin{equation}
\sum_{i=1}^{j-1} V_{M;i} < V_n \leq \sum_{i=1}^{j} V_{M;i},
\end{equation}
and
\begin{eqnarray}
V_n &\!\!\!=\!\!\!& \sum_{i=1}^{j-1} V_{M;i} + \nonumber\\
&&\frac{8\pi^2}{3} (r_{b;j}^3 \!-\! r_{a;j}^3) 
(v_{r,b;j}^{} \!-\! v_{r,a;j}^{})(v_{t,b;j}^2 \!-\! v_{t,a;j}^2) f_{n}^{}.
\end{eqnarray}
Then, in cell $j$ the co-ordinates $r_{a;j}^3 \leq r_{n}^3 \leq r_{b;j}^3 $\,,
$v_{r,a;j} \leq v_{r,n} \leq v_{r,b;j}$ and $v_{t,a;j}^2 \leq v_{t,n}^2
\leq v_{t,b;j}^2$ are randomly generated. Thus, a point 
$(r_n,v_{r,n},v_{t,n},f_n)$ is uniformly chosen
in the 7-dimensional phase-space volume $V_M$. Now, if $f_n \leq F(r_n,v_{r,n},v_{t,n})$, the
co-ordinate $(r_n,v_{r,n},v_{t,n})$ is accepted as a valid data point,
otherwise it is rejected. Furthermore, if $f_{b;j} < F(r_n,v_{r,n},v_{t,n})$, the cell
volume is accordingly increased to the new maximum, so the grid keeps
being improved. 

In this manner we construct a data set of $N$ accepted co-ordinates
inside the chosen radius $r_b$ which follow the distribution. The MC
simulation is regarded successful if the cell volumes have changed
negligibly (if the relative change of the total volume is smaller than
$10^{-3}$) during the MC simulation. If not, a new MC simulation with
the final grid (with volume $V_{M+N}$) is necessary.  Also, if the
ratio between rejected and accepted points is very large, causing the
MC simulation to be slow, the grid might have to be refined further
(as aforementioned, we stop refining the grid once the cells cover a
volume that is a factor 1.5 to 5 larger than the model's actual
phase-space volume).

Finally, every co-ordinate $(r_n,v_{r,n},v_{t,n})$ has to be converted
into a phase-space point $(x_n,y_n,z_n,v_{x,n},v_{y,n},v_{z,n})$. This
is done by uniformly simulating the surface of a sphere with radius
$r_n$ (creating $(x_n,y_n,z_n)$), a circle with radius $v_{t,n}$
(creating $(v_{\theta,n},v_{\phi,n})$) and the sign of $v_{r,n}$.  The
velocities can then be transformed into the appropriate Cartesian
co-ordinates. For isotropic functions $F(E)$ the grid abscissae simplify to the
2-dimensional $(r,v)$ space, and the entire procedure is
analogous. 

Our method has several advantages: the construction of a grid and the
subsequent MC simulation of points is straightforward, fast, accurate
and generally applicable. This contrasts with algorithms that require
integrations and inversions of DFs, which can experience numerical
problems with intricate functions. Also, no intermediate steps are
required (e.g.\ simulating the density first before assigning velocities
to each particle) and once a grid is made for a model,
it can be re-used to generate an arbitrary number of particles.
Moreover, since a peak can be
adequately isolated by a cell, infinite ranges in the co-ordinate
space or the DF values can be approximated by choosing appropriate
large boundary values.

As an example we show in Fig.~\ref{Figgrid} the constructed cells for
an isotropic Hernquist system with a central SBH of $\mu=0.1$ (see
Section 4). A simulated data sample ($10^5$ accepted particles) for an
anisotropic Hernquist system with a central SBH of $\mu=0.1$ and an
anisotropy radius of $r_a=1$ is shown in Fig.~\ref{Figsimulatie}.  In
all our MC simulations, we truncate the infinite boundary radius at
$r_b=2000$. For the DFs with an infinite maximum, we set $f_b=10^{15}$,
 and for the SBH-models we set the maximum velocity at the
arbitrarily large value $v_b=10^{15}$ (these values are in fact much larger than
needed. In reality, no particle is ever
assigned such a high DF value or initial velocity and never reaches such high
velocities during the subsequent $N$-body simulations).

\subsection{$N$-body code}
We studied the stability of our models by using an $N$-body code that
is based on the ``self-consistent field'' method \citep{hern2}. This
method relies on the series expansion in a bi-orthogonal spherical
basis set for both the density and gravitational potential
\begin{eqnarray}
\rho (r,\theta ,\phi)&=&\sum_{nlm}A_{nlm}\ \rho_{nlm}(r,\theta
,\phi)\nonumber\\&=&\sum_{nlm}A_{nlm}\ \tilde{\rho}_{nl}(r)\ Y_{lm}(\theta ,\phi),
\end{eqnarray}
\begin{eqnarray}
\Phi (r,\theta ,\phi)&=&\sum_{nlm}A_{nlm}\ \Phi_{nlm}(r,\theta
,\phi)\nonumber\\&=&\sum_{nlm}A_{nlm}\ \tilde{\Phi}_{nl}(r)\ Y_{lm}(\theta ,\phi),
\label{pot}
\end{eqnarray}
where $Y_{lm}(\theta ,\phi)$ are the spherical harmonics. Some freedom
is considered for this expansion since
($\tilde{\rho}_{nl}(r)$,$\tilde{\Phi}_{nl}(r)$) can have different
forms (e.g.\ Plummer model, Bessel functions, spherical harmonic
functions), however here we will use a form similar to the Hernquist
model due to its trivial connection with our anisotropic systems that
we wish to examine:
\begin{eqnarray}
\tilde{\rho}_{nl}(r)&=&\frac{K_{nl}}{\sqrt{\pi}}\frac{r^l}{r(1+r)^{2l+3}}C_n^{(2l+3/2)}(\xi),\label{nbodyeq1}\\
\tilde{\Phi}_{nl}(r)&=&-2\sqrt{\pi}\frac{r^l}{(1+r)^{2l+1}}C_n^{(2l+3/2)}(\xi),
\label{nbodyeq2}
\end{eqnarray}
where $K_{nl}$ is a normalization constant, $\xi=(r-1)/(r+1)$
and $C_n^{(2l+3/2)}(\xi)$ are Gegenbauer polynomials
(e.g. Szeg\"o (1939), Sommerfeld (1964)). The coefficients
$A_{nlm}$ can be calculated by means of all the particles that
describe the DF of our system (see Hernquist \& Ostriker (1992) for
more details). The spherical accelerations for each particle are found
by taking the gradient of the potential (eq. \ref{pot}). Finally new
positions and velocities are derived with the use of an integrator
which is equivalent to the standard time-centred leapfrog
\citep{allen,hut},
\begin{eqnarray}
x_{i+1}&=&x_i+\Delta tv_i+\frac{1}{2}\Delta t^2a_i,\\
v_{i+1}&=&v_i+\frac{1}{2}\Delta t(a_i+a_{i+1}).
\end{eqnarray}
The indices $n$, ($l$, $m=-l...l$) are indirectly an indication for
the accuracy of the simulation for respectively radial and tangential
motion, since they determine the number of terms in the expansion (see
Section~5.2 Hernquist \& Ostriker (1992) for a statistical analysis).
For the systems without the SBH we find that $n_{max}$=4 and
$l_{max}$=2 assures a total energy conservation of better than $\sim
10^{-6}$ over 50 half-mass dynamical times $T_h$ and still allows a
low CPU time per $N$-body time-step (the time between two successive
calculations) of $\Delta t=T_h/416 \approx 0.02$. For the systems with
an SBH, we used $n_{max}=6,l_{max}=2$ when $\mu \leq 0.05$, and
$n_{max}=8,l_{max}=4$ for larger values of $\mu$. The gravitational
effect of the SBH is added analytically by an extra radial
acceleration proportional to the mass of the SBH. To avoid numerical
divergences when particles pass close to the SBH, we included a
softening to this acceleration, i.e. $-\mu/(r^2+\epsilon^2)$ with
softening length $\epsilon=0.05$. At this radius the dynamical
crossing time of a particle is $T_h=0.37$, which is still notably
larger than our time-step of $0.02$. Adding this softening causes a
discrepancy in the treatment of the SBH potential between the
analytical models and the N-body code. As a consequence the initial
conditions are not exactly in dynamical equilibrium so that the
systems develop transient radial motions to adjust their density
profile. However, this effect is marginal and does not influence the
overall results. In all simulations with an SBH the energy is
conserved better than 1\% over 50 half-mass dynamical times.

In order to check the robustness of our results, we performed two kinds
of tests. We {\em (i)} re-ran a number of simulations with different,
smaller time-steps, and {\em (ii)} we performed simulations with higher
$n_{max}$ and $l_{max}$ values. A detailed comparison of these extra runs
with the original simulations shows that our results and conclusions
do not change : the variation of the global instability indicators,
such as axis ratios or $2 K_r/K_t$, as a function of time are
essentially the same.

\subsection{Quantifying the instabilities} \label{axesstab}
When a system is unstable, it tends to create a bar feature at its
center (see Fig. \ref{bar}) which roughly has an ellipsoidal shape. As
noted by other authors \citep{merritt,papa}, the physical cause of
instability is similar to that of the formation of a bar in a disc
\citep{lynden}, where a small perturbation changes the orbits with a
lower angular momentum (initially precessing ellipses) into boxes
which are aligned along the initiated bar. A particle in a box orbit
is unable to precess all the way round and will fall each time back to
the bar. This effect will cause the bar to increase in both size and
strength. To measure the radial stability of the systems we fitted the
shape of an ellipsoidal mass distribution by means of an iterative
procedure \citep{dub,katz,meza1,meza2} at every half-mass dynamical
time.  This detects any bar feature that is located within a
given radius. The initial condition of this method is
\begin{equation}
\rho=\rho(a)\ \ \mathrm{with}\ \
a=\left(x^2+\frac{y^2}{q^2}+\frac{z^2}{s^2}\right)^{1/2},
\nonumber
\end{equation}
and with $M_{ij}=\sum\frac{x_ix_j}{a^2}$, the principal components of
the inertia tensor $M_{zz}\le M_{yy}\le M_{xx}$ and the axis ratios
$q$ and $s$ equal to 1, assuming a spherical mass distribution within
a certain sphere with a given radius for which we chose $r=5$. For
all considered models this radius encloses approximately 70\% of the
total mass.  To achieve these conditions a transition to the center of
mass has to be made followed by swapping the coordinate axes into the
correct order. In the next step the eigenvalues and eigenvectors of
the inertia tensor $I_{ij}$ are calculated, transforming it into a
diagonal matrix. At this point the new axis ratios can be calculated
\begin{equation}
q=\left(\frac{M_{yy}}{M_{xx}}\right)^{1/2}\!\!\!=
\frac{b}{a}\ \ \mathrm{and}\ \ s=\left(\frac{M_{zz}}{M_{xx}}\right)^{1/2}\!\!\!=\frac{c}{a},
\end{equation}
which in turn are used as the conditions for the next iteration
step. The iteration was stopped as soon as both axis ratios converged
to a value within a pre-established tolerance of $10^{-3}$. Thus at
each half-mass dynamical time the values of these axis ratios
serve as measures of the strength of the bar instability, if
present.

\section{Hernquist models without a black hole}
In this section we investigate the stability of two different families
of anisotropic Hernquist models without a central supermassive black
hole.  For the analytical construction of these models we refer to
Baes \& Dejonghe (2002), however we will recapitulate the
characteristics of each family.
\begin{figure}
\centering
\includegraphics[width=8cm,clip]{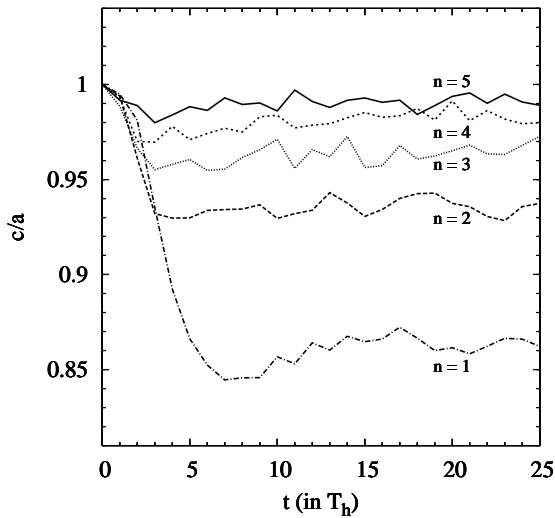}
    \caption{Axis ratios as a function of time of the anisotropic
    systems with a decreasing anisotropy.}
    \label{decraxis}
\end{figure}
\subsection{Family I: Decreasing anisotropy}
We find Hernquist models with a decreasing anisotropy by assuming an
augmented density of the form
\begin{equation}
\tilde{\rho}(\psi
,r)=\frac{1}{2\pi}\frac{\psi^{4-2\beta_{n}}}{(1-\psi)^{1-2\beta_0}}
\frac{(1+r)^{2(\beta_0-\beta_{n})}}{r^{2\beta_0}},
\end{equation}
with $\beta_{n}=\beta_0-\frac{n}{2}$ and $n$ a natural
number. After some algebra we find the distribution function
\begin{eqnarray}
F(E,L)\rev&=&\rev\frac{2^{\beta_0}}{(2\pi)^{5/2}}\Gamma(5-2\beta_{n})L^{-2\beta_0}
E^{5/2-2\beta_{n}+\beta_0} \nonumber\\ 
\rev&\times&\rev \sum_{k=0}^n\!{n \choose k}
\!\frac{1}{\Gamma(\frac{2+k}{2}\!-\!\beta_0)\Gamma(\frac{7-k}{2}     
\!-\!2\beta_{n}\!+\!\beta_0)}\!\!\left(\!\frac{L}{\sqrt{2E}}\!\right)^k\nonumber\\
\rev&\times&\rev
_2F_1\left(5\!-\!2\beta_{n},1\!-\!2\beta_0;
\frac{7-k}{2} 
\!-\!2\beta_{n}\!+\!\beta_0;E\right),
\end{eqnarray}
with $_2F_1$ a hypergeometric function (see Appendix A), and the
anisotropy
\begin{equation}
\beta(r)=1-\frac{\sigma^2_t(r)}{\sigma_r^2(r)}=\frac{\beta_0+\beta_{n}r}{1+r},
\end{equation}
which decreases as a function of radius. Since for $\beta_0\le 0$ we
only find tangentially dominated systems which are free of radial
instabilities, we limit ourselves to the investigation of the case
$\beta_0=0.5$. For this value, $n=0$ corresponds to a system with
constant anisotropy. We plotted the axis ratios $c/a$ for a number of
different models with different $n$ in Fig.~\ref{decraxis}. Here and
in the remainder of the paper, we define those models that keep the
axis ratio $c/a \gtrsim 0.95$ over 50 dynamical times as being
stable. The only model that does not satisfy this criterion is that
with $n=1$, which is everywhere radially anisotropic. For $n \ge 2$,
the models become tangentially anisotropic at larger radii and as a
consequence are much more stable. This is evident from
Fig.~\ref{decraxis}. It is clear that the minimum of $c/a$ is reached
rapidly, whereafter the systems are in an equilibrium state, but are
slightly non-spherical.

\begin{figure}
\centering
\includegraphics[width=8cm,clip]{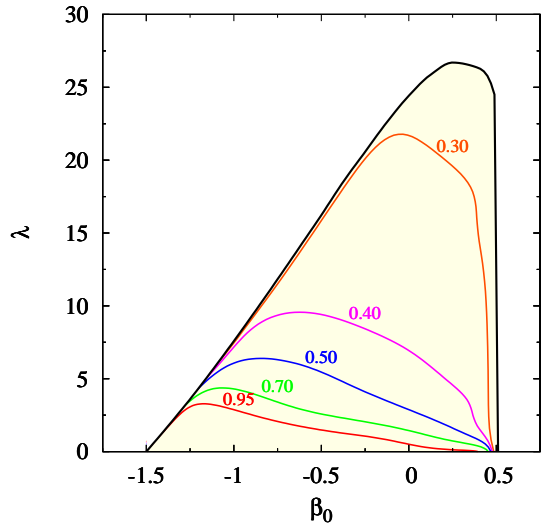}
    \caption{The stability of the Hernquist models without a black
      hole and with increasing anisotropy (see Section
      \ref{sectionincmod}), expressed as the minima of the axis ratios
      $c/a$ during the simulations.  The shaded area indicates the
      region of physical systems, i.e.\ with a non-negative
      distribution function.}
    \label{incrani}
\end{figure}
\begin{figure*}
\centering
\includegraphics[width=18cm,clip]{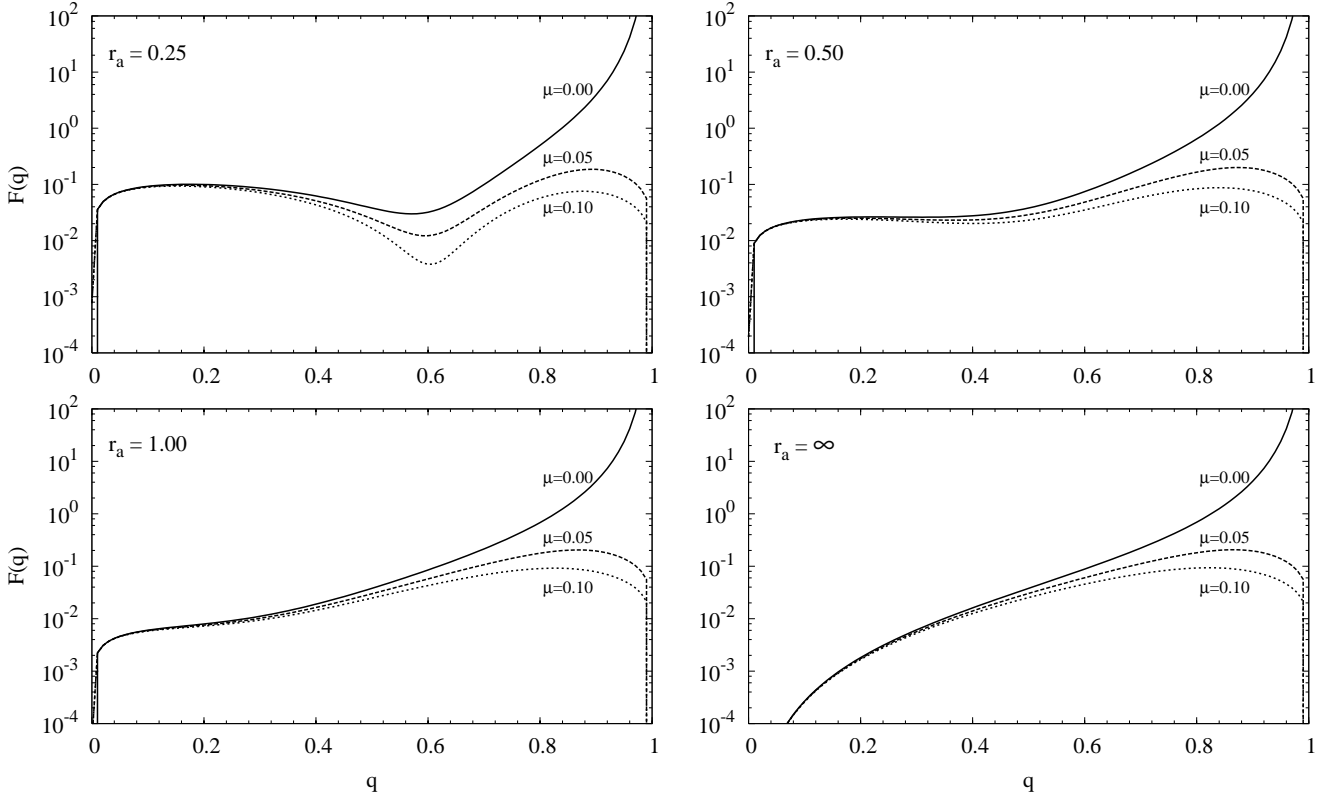}
     \caption{The distribution functions defined by Ciotti (1996),
       with different anisotropy radii $r_a$ and SBH mass $\mu$. The 
       value $r_a=\infty$ corresponds with the isotropic case;
       for $\mu=0$ the DFs reduce to eq.~(\ref{hernb0}).}
     \label{SBHdfs}
\end{figure*}
\subsection{Family II: Increasing anisotropy}
\label{sectionincmod}
These models are a generalization of the Osipkov-Merritt models
\citep{cudd} with an augmented density and DF of the general form
\begin{eqnarray}
\tilde{\rho}(\psi,r)\rev&=&\rev r^{-2\beta_0}f(\psi)\left(1+\lambda r^2\right)^{-1+\beta_0}
\ \ \textrm{with}\ \ \lambda=\frac{1}{r_a^2},\\
F(E,L)\rev&=&\rev F_0(Q)L^{-2\beta_0}\ \ \textrm{with}\ \ 0 \leq Q=E-\frac{L^2}{2r_a^2} \leq 1,
\label{Q}
\end{eqnarray}
and $E$ denotes the energy, $L$ the angular momentum and $r_a$ the
anisotropy radius. The explicit form of $f(\psi)$ for the Hernquist
potential-density pair can be found in Baes \& Dejonghe (2002).  As
mentioned by them, the DFs can be written analytically for the
half-integer values $\beta_0=0.5, 0, -0.5, -1$, so we will limit
ourselves to these cases. For every value of $\beta_0$, we also
computed numerically the maximum anisotropy value
$\lambda_\mathrm{max}(\beta_0)$, outside which the DFs become negative
for some values of $Q$ and $L$.  The area of physical systems is
indicated in Fig.~\ref{incrani}. Our models have the following
functional form:
\begin{itemize}
\renewcommand{\labelitemi}{$\bullet$}
\setlength{\listparindent}{0cm}
\item{$\beta_0=0.5$:
\begin{equation}
F(E,L)=\frac{Q}{4\pi^3L}\frac{3Q^2+\lambda(3Q^2-5Q+2)}{\sqrt{Q^2+\lambda(1-Q)^2}}.
\end{equation}}
\item{$\beta_0=0$:
\begin{eqnarray}
F(E,L)\rev&=&\rev\frac{1}{8\sqrt{2}\pi^3}\left[\frac{3\arcsin{\sqrt{Q}}}{(1-Q)^{5/2}}\right. 
\nonumber \label{hernb0}\\
\rev&&\rev+\left. \sqrt{Q}(1-2Q)\left(\frac{8Q^2-8Q-3}{(1-Q)^2}+8\lambda\right)\;\right].
\end{eqnarray}}
\item{$\beta_0=-0.5$:
\begin{eqnarray}
F(E,L)=\frac{L f(Q)}{4\pi^3(1-Q)^4\sqrt{Q^2+\lambda(1-Q)^2}},
\end{eqnarray}
with
\begin{eqnarray}
f(Q)&=&6(1+\lambda)^2Q^6-2(16\lambda^2+26\lambda+10)Q^5\nonumber\\&&
+(70\lambda^2+87\lambda+20)Q^4-2\lambda(40\lambda+33)Q^3\nonumber\\&&
+\lambda(50\lambda+19)Q^2-16\lambda^2Q+2\lambda^2.
\end{eqnarray}}

\item{$\beta_0=-1$: 
\begin{eqnarray}
F(E,L)&=&\frac{L^2}{256\sqrt{2}\pi^3(1-Q)^5} \times\\&& 
\left[\frac{f_1(Q)}{\sqrt{1-Q}}
\arctan\left(\frac{\sqrt{Q}}{\sqrt{1-Q}}\right) + \frac{f_2(Q)}{\sqrt{Q}}\right],
\end{eqnarray}
with
\begin{equation}
f_1(Q)=15\left[(16\lambda +120)Q^2-(72+32\lambda )Q+15+16\lambda\right],
\end{equation}
\begin{eqnarray}
f_2(Q)&=&384(1+\lambda)^2Q^6\nonumber\\&&
-(1984\lambda^2+3712\lambda+1728)Q^5\nonumber\\&&
+(4160\lambda^2+7008\lambda+2784)Q^4\nonumber\\&&
-(4480\lambda^2+6192\lambda+1200)Q^3\nonumber\\&&
+(2560\lambda^2+2368\lambda+930)Q^2\nonumber\\&&
-(704\lambda^2+240\lambda+225)Q+64\lambda^2.
\end{eqnarray}}
\end{itemize}
For all models the anisotropy is given by the simple formula
\begin{equation}
\beta(r)=\frac{r^2+\beta_0r_a^2}{r^2+r_a^2},
\end{equation}
showing an increase in anisotropy as a function of radius. The results
of the $N$-body investigation for all $\beta_0$ and $\lambda$ are
summarized in figure \ref{incrani}, where we plotted the minimal axis
ratios $c/a$ for the DFs in this parameter space. To derive this plot,
we simulated systems with $\beta_0= 0.5, 0, -0.5, -1$ and
$\lambda=1,2,4,6,10,16,24$ where physically possible.  The case where
$\beta_0=0$ corresponds to the traditional anisotropic Osipkov-Merritt
Hernquist model that has been previously investigated in a similar way
by Meza \& Zamorano (1997). These authors state the system with $r_a
\approx 1.1$ (or $\lambda \approx 0.82$) as stable. To compare our
study with theirs, we simulated this model in addition to the other
systems. For this model we find an axis ratio $c/a \approx 0.95$ after
50 dynamical times, and $2 K_r/K_t \approx 2.2$ during the entire
run. These values are in agreement with their results, therefore we
will define $c/a = 0.95$ as our stability criterion.

As is to be expected, the
anisotropy radius $\lambda$ strongly affects the formation of
radial-orbit instabilities, so that only models with a low value of
$\lambda$ remain stable.  Furthermore we note that all models remain
in their new equilibrium state after $t\approx 10T_h$, as in case of
the DFs of Family I. As an example, the $c/a$ ratio evolution for one
of the systems is given in Fig.~{\ref{bar}}.

\begin{figure}
\centering
\includegraphics[width=8cm,clip]{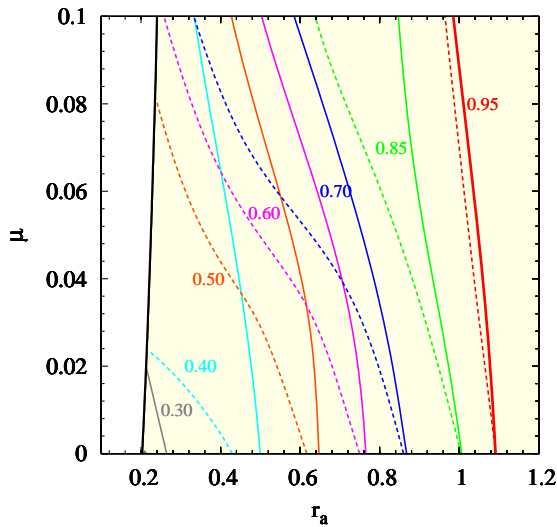}
\caption{ Contour plot of the axis ratio $c/a$ for the
Osipkov-Merritt models as a function of anisotropy radius and mass of
the SBH. The shaded area indicates the region of physical systems,
i.e.\ with a non-negative distribution function. The solid lines
indicate the minimal values during the simulation, the dashed lines
show the axis ratios at the end of the simulation (at $t=50T_h$).}
    \label{contourc}
\end{figure}
\section{Hernquist models with a supermassive black hole}
The now established presence of diverse components in a great variety
of galaxies calls for more advanced dynamical models. In this respect
a dark matter halo and a central supermassive black hole are important
and can change the galaxy's properties dramatically. However, up to
now there are few known analytical systems that include e.g.\ a
supermassive black hole. The only models known so far are presented in
Ciotti (1996), Baes \& Dejonghe (2004) and Baes et al.\ (2005) which
are all based on the $\gamma$-models with special attention to the
Hernquist model and in Stiavelli (1998) where the distribution
function of a stellar system around an SBH is derived from statistical
mechanic considerations.

In this section we investigate the radial stability of both isotropic
and anisotropic Hernquist models containing a supermassive black hole,
as these represent the closest analytical approach to the
observations. For the following sections we will use the
representation of Ciotti (1996); again, we are not going into great
detail in the derivation of the analytical distribution function.

In essence the DFs are obtained from an analytical Osipkov-Merritt
inversion of the systems governed by eq.~(\ref{hernpsi}) and
(\ref{hernrho}). As a consequence, these models can be viewed as a
extension of eq.~(\ref{hernb0}). Subsequently, we will refer to
these combined systems as Osipkov-Merritt models.
The DFs can be written as
\begin{equation}
F(Q)=F_i(Q)+\frac{F_a(Q)}{r_a^2},
\label{hoofd}
\end{equation}
where $Q$ has the same definition as in eq.~(\ref{Q}). A more 
natural parameter $q$ is defined through
\begin{equation}
Q=q\left(1+\frac{\mu}{1-q}\right),\quad 0 \leq q \leq 1.
\end{equation}
\begin{figure*}
\centering
\includegraphics[width=18cm,angle=0]{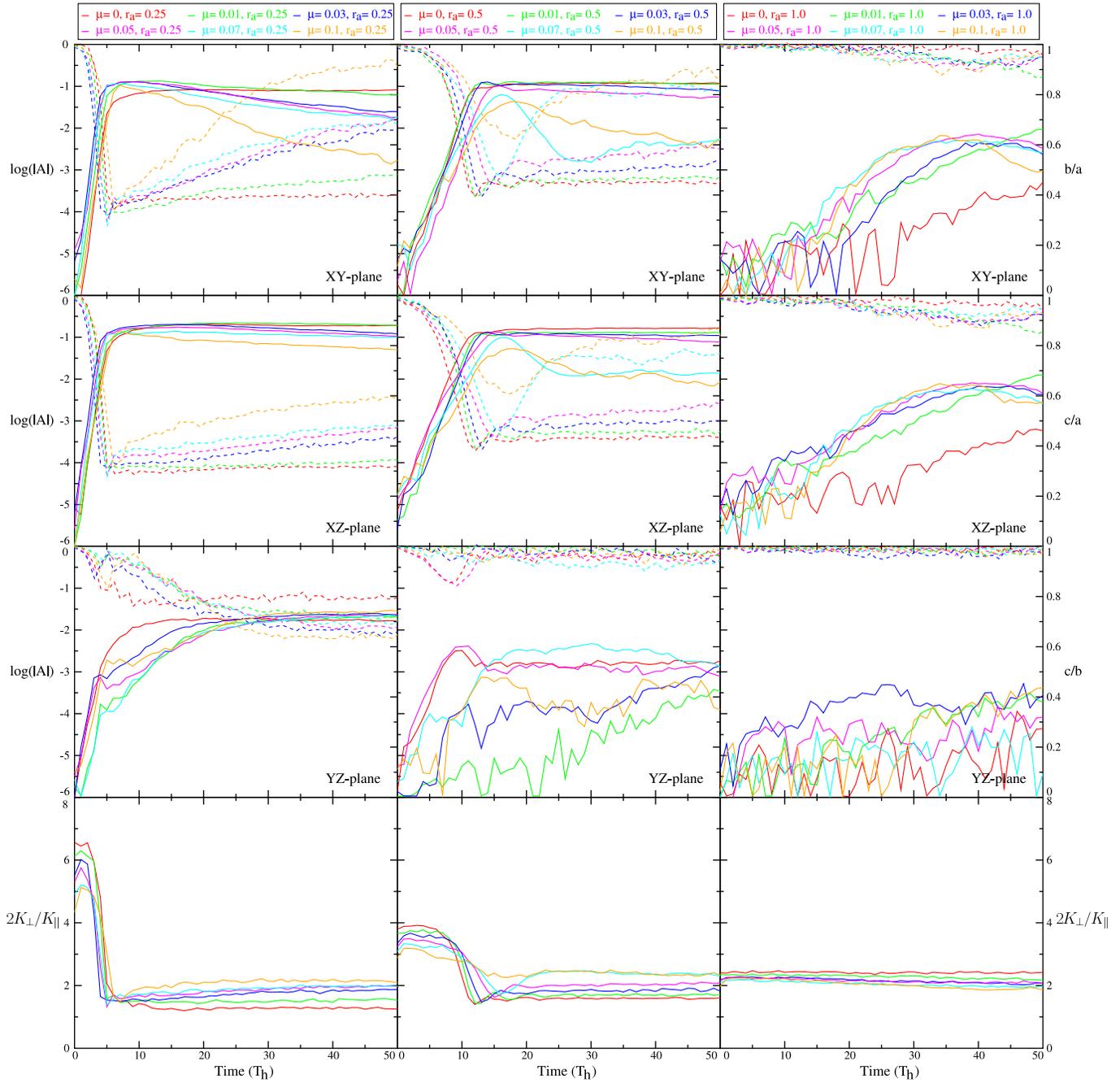}
\caption{The evolution of systems with an anisotropy radius of
respectively $r_a=0.25, 0.50, 1.0$, but with different mass of the SBH
($\mu=0, 0.01, 0.03, 0.05, 0.07, 0.1$). For each column we plot the
evolution of the axes ratios (dashed lines) and Fourier coefficients
(full lines) in the upper three rows. The bottom row contains the
evolution of the $2K_{\perp}/K_{\parallel}$ ratio as a function of dynamical
time ($T_h$).}
    \label{krkt}
\end{figure*}
\subsection{Family I: Isotropic}
We find an isotropic system by letting $r_a$ diverge to $\infty$. Then
eq.~(\ref{hoofd}) simplifies to
\begin{equation}
F(E)=F_i(E)=\frac{1}{2\sqrt{8}\pi^3}\left(\frac{dE}{dl}\right)^{-1}
\frac{d}{dl}\left[\tilde{F}_i^{\pm}(l)\right],
\end{equation}
with the argument $l$ defined as $l^2=1-q$.
For $\tilde{F}_i^{\pm}(l)$ we refer to
Ciotti (1996) as this involves combinations of elliptic and Jacobian
functions. These models only differ from those of Baes et
al.\ (2005) in the definition of the parameter $\mu$.

Although the systems are isotropic, their DFs have a local maximum
when $\mu > 0$ (as shown in Fig.~\ref{SBHdfs}). Hence, the sufficient
criteria of Antonov (1962) and Dor\'emus and Feix (1973) for isotropic
systems cannot be applied. However, in our subsequent analysis of the
systems with and without an SBH in Section 5, it will be shown that
all models with $r_a>1$ are stable. In other words, it becomes evident
that the addition of a central SBH, although it changes the dynamics
dramatically, does not influence the density distribution of an
isotropic system.

\subsection{Family II: Anisotropic}
In a similar way as the isotropic case the distribution function can be written as
\begin{eqnarray}
F(Q)&=&F_i(Q)+\frac{F_a(Q)}{r_a^2},\\
&=&\frac{1}{2\sqrt{8}\pi^3}\left(\frac{dQ}{dl}\right)^{-1}\frac{d}{dl}
\left[\tilde{F}_i^{\pm}(l)+\frac{\tilde{F}_a^{\pm}(l)}{r_a^2}\right],
\end{eqnarray}
where again $\tilde{F}_a^{\pm}(l)$ is defined in Ciotti (1996).  In
Fig.~\ref{SBHdfs} we display systems with several values of $\mu$ and
$r_a$.  Notice that for small values of $r_a$ the DFs have a local
minimum. As a consequence, for every $\mu$ there exists a smallest
possible $r_a$, where this minimum becomes zero; smaller values
of this boundary $r_a$ result in negative DFs, thus creating
unphysical systems. For $\mu=0$, the minimal anisotropy radius
is $r_a\approx 0.202$; for $\mu=0.1$ the boundary becomes $r_a\approx 0.240$. From
the viewpoint of a stability analysis these systems are the most
interesting. In the following section we will discuss their evolution
in detail, comparing them with the models without an SBH
(eq.~(\ref{hernb0})).

\section{Stability analysis of the Osipkov-Merritt models}

\begin{figure*}
\centering
\includegraphics[width=18cm,clip]{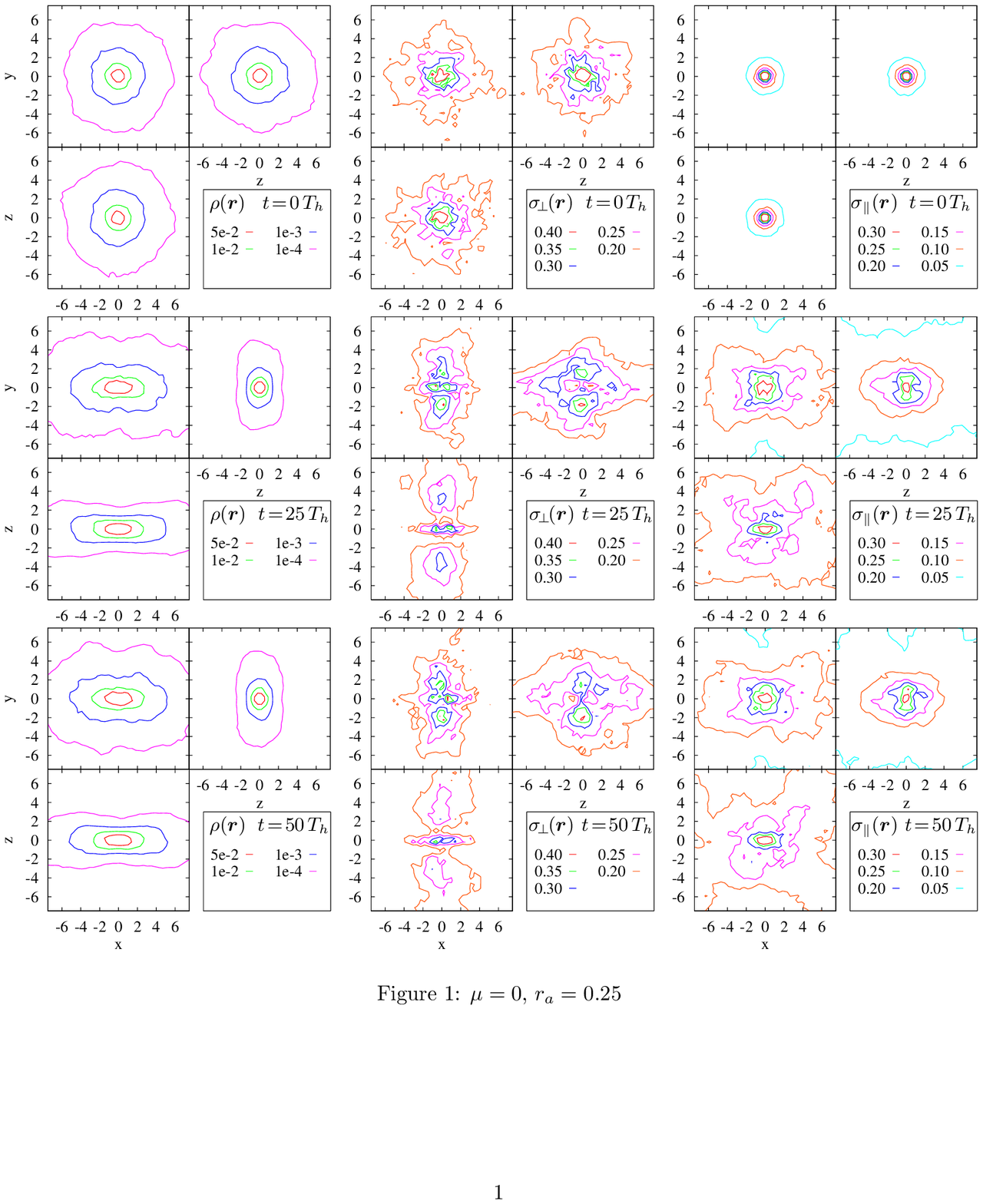}
\caption{The spatial density $\rho$ and the velocity dispersions
$\sigma_{\!\perp}$ and $\sigma_\parallel$, in the three principal
planes, of an Osipkov-Merritt Hernquist system without an SBH, and
with $r_a=0.25$. Dynamical times $t=0$, $t=25T_h$ and $t=50T_h$ are
displayed.}
    \label{betasrho}
\end{figure*}
\begin{figure*}
\centering
\includegraphics[width=18cm,clip]{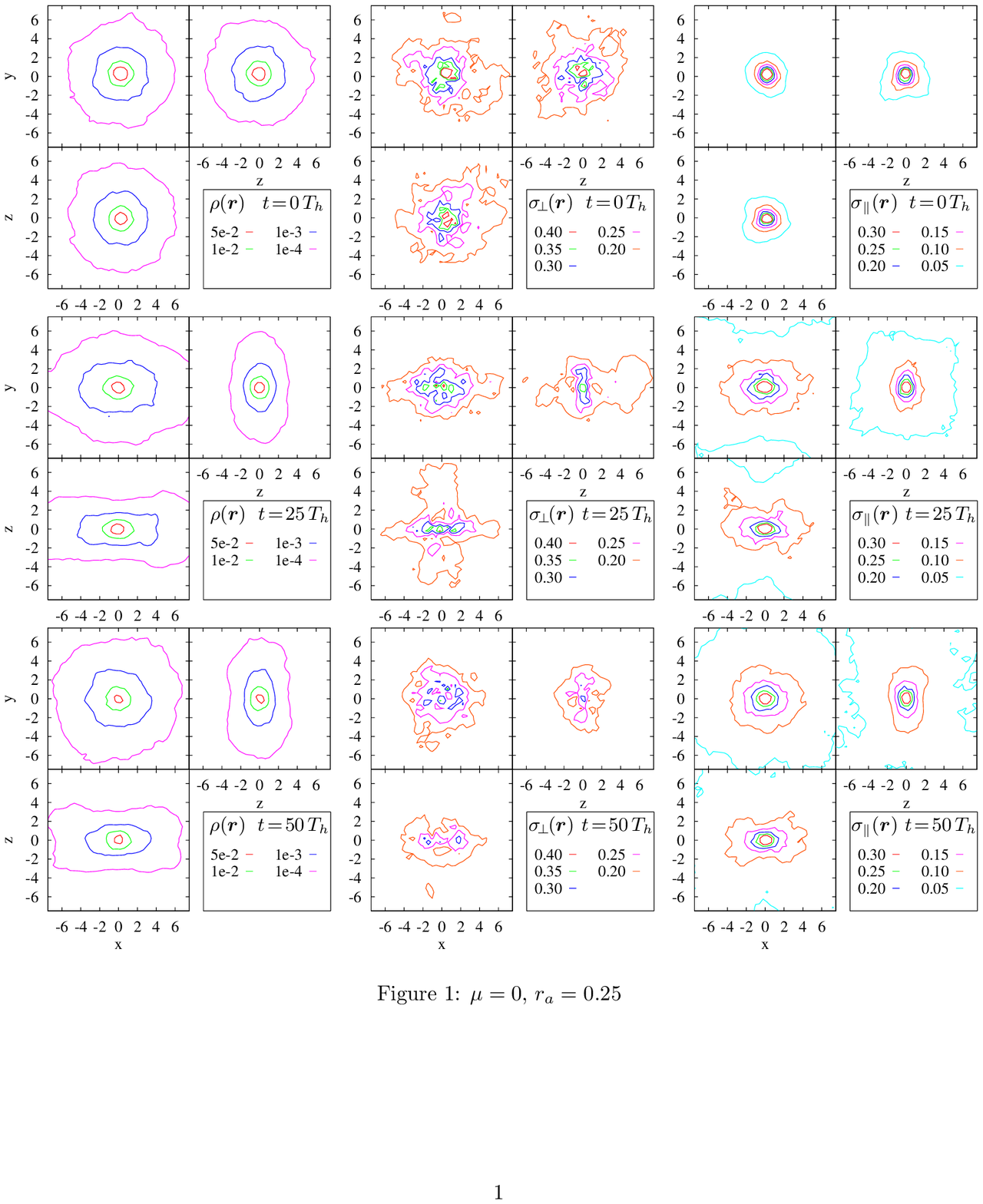}
\caption{The spatial density $\rho$ and the velocity dispersions
$\sigma_{\!\perp}$ and $\sigma_\parallel$, in the three principal
planes, of an Osipkov-Merritt Hernquist system with an SBH of
$\mu=0.05$, and with $r_a=0.25$. Dynamical times $t=0$, $t=25T_h$ and
$t=50T_h$ are displayed.}
    \label{betasrho2}
\end{figure*}
\begin{figure*}
\centering
\includegraphics[width=18cm,clip]{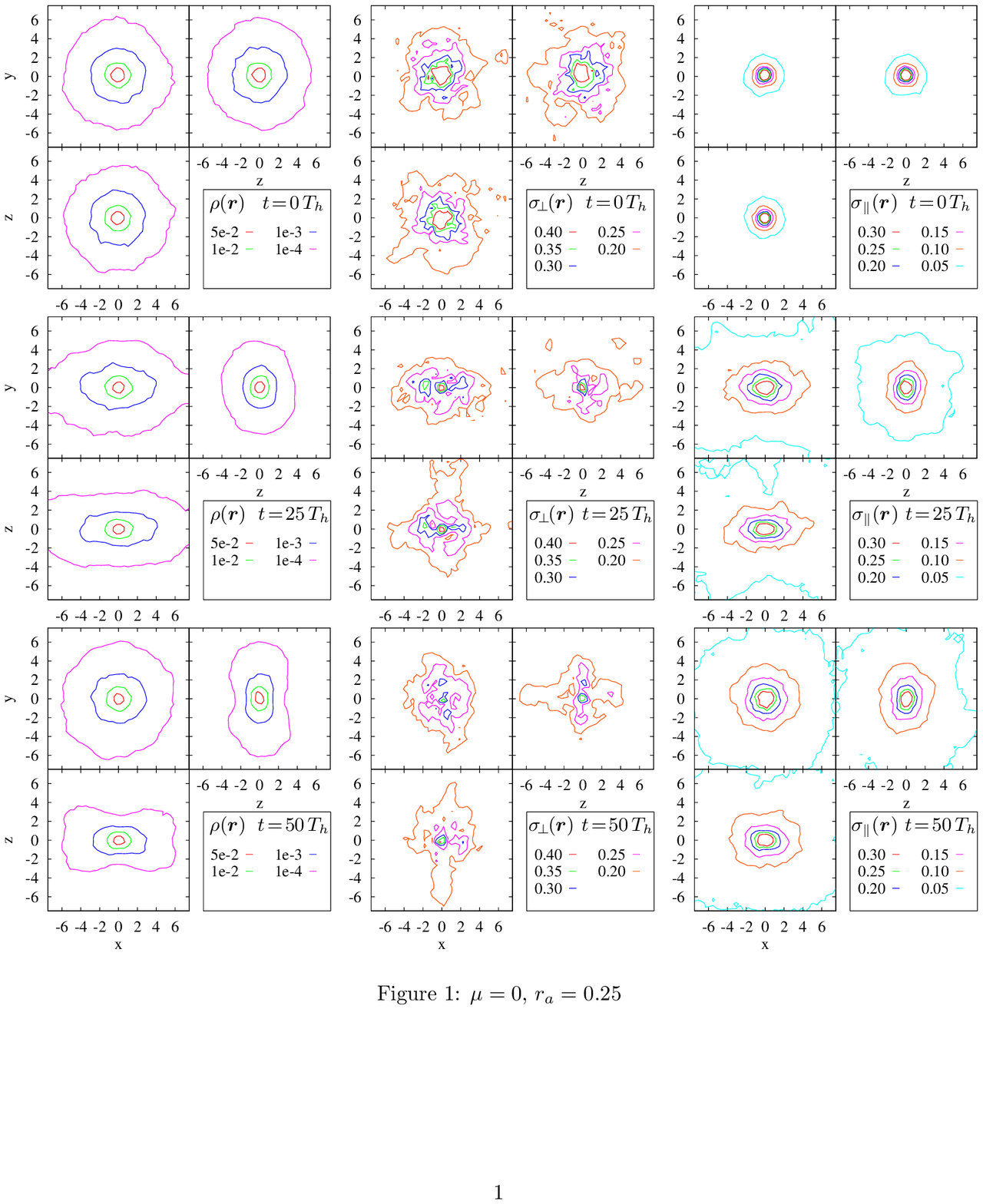}
\caption{The spatial density $\rho$ and the velocity dispersions
$\sigma_{\!\perp}$ and $\sigma_\parallel$, in the three principal
planes, of an Osipkov-Merritt Hernquist system with an SBH of
$\mu=0.1$, and with $r_a=0.25$. Dynamical times $t=0$, $t=25T_h$ and
$t=50T_h$ are displayed.}
    \label{betasrho3}
\end{figure*}
\begin{figure*}
\centering
\includegraphics[width=18cm,clip]{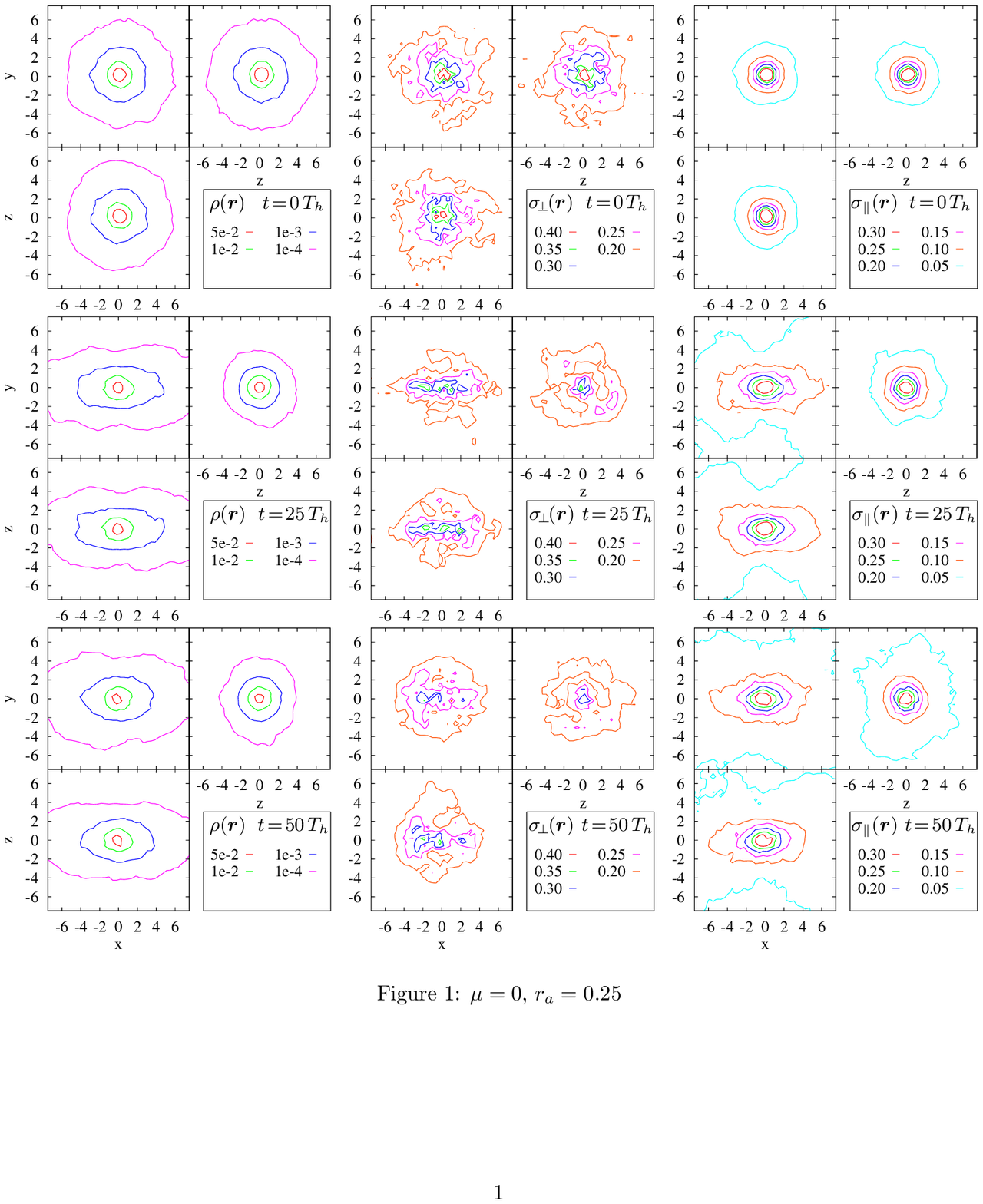}
\caption{The spatial density $\rho$ and the velocity dispersions
$\sigma_{\!\perp}$ and $\sigma_\parallel$, in the three principal
planes, of an Osipkov-Merritt Hernquist system with an SBH of
$\mu=0.05$, and with $r_a=0.50$. Dynamical times $t=0$, $t=25T_h$ and
$t=50T_h$ are displayed.}
    \label{betasrho4}
\end{figure*}

To investigate any trend about the radial stability of these systems,
we investigate the 2-parameter space ($r_a$,$\mu$). In total, we
performed 25 simulations, with $r_a=0.25,0.50,0.70,0.85,1.00$ and
$\mu=0.01,0.03,0.05,0.07,0.10$. This way we derived a grid of values
of the $c/a$ axis ratios, shown in Fig.~\ref{contourc}. As stated
before, the ($\mu=0$)-axis corresponds to the systems in
eq. (\ref{hernb0}).  The solid lines indicate the minimal values
reached during the simulation (i.e.\ when the instability is
strongest). The rate at which these minima are reached strongly
depends on $r_a$, ranging from a few half-mass dynamical times for
highly radial models to $t\approx 50T_h$ for systems with $r_a=1.00$.
After the point of time upon which a system obtains its minimal $c/a$
the influence of the SBH causes a diminution of the bar instability,
resulting in the $c/a$ axis ratios at $t=50T_h$ shown by the dashed
lines. Thus, in each system the particles are affected by two
counteracting forces: the (relatively fast) bar formation and the
(more gradually) scattering near the center due to the spherically
symmetric gravitational potential of the black hole. The contour line
$c/a=0.95$ is highlighted as our stability criterion. 

A full dynamical analysis would require a detailed study of the
orbital distribution of the stellar mass. However, we can gain
important insights into the dynamics of the models by visualizing
the evolution of various quantities, in 
Figs.~\ref{krkt}-\ref{betasrho4}.

First, in order to retain a notion of 'radial' and 'tangential' motion
in an evolved system (resembling a triaxial model) at a certain time
$t$ , we use the method described in Section~\ref{axesstab} to
approximate the mass distribution inside the radius $r_i$ of each
particle by an ellipsoid. Then, the velocity of a particle can be
written into two components perpendicular resp.\ parallel to its
surface $v_i = v_{i,\perp} + v_{i,\parallel}$. Subsequently, the
perpendicular and parallel velocity dispersion of the $m$ nearest
neighbors around a position $\bmath{r}$ are
\begin{eqnarray}
\sigma^2_{\perp}(\bmath{r}) &=& \frac{1}{m-1}\sum_{i=1}^m\big(v_{i,\perp}
-\bar{v}_{\perp}\big)^2,\\
\sigma^2_{\parallel}(\bmath{r}) &=& \frac{1}{2(m-1)}\sum_{i=1}^m\big(v_{i,\parallel}
-\bar{v}_{\parallel}\big)^2.
\end{eqnarray}
In a similar manner we define
\begin{eqnarray} 
K_{\perp} &=& \frac{1}{N}\sum_{i=1}^N v_{i,\perp}^2,\\
K_{\parallel} &=& \frac{1}{N}\sum_{i=1}^N v_{i,\parallel}^2,
\end{eqnarray}
so that $2K_{\perp}/K_{\parallel}$ can serve as a non-spherical extension
of $2K_r/K_t$.
 
In Fig.~\ref{krkt} we compared our $c/a$
criterion to other known methods to quantify the stability of the
system. For all our systems with $r_a$ = 0.25, 0.50 and 1.0,
we plot the evolution of the three axis
ratios against the Fourier coefficient \citep{sellwood94}
in the three principal planes (XY, XZ and YZ), and the evolution of 
$2K_{\perp}/K_{\parallel}$(t). The Fourier coefficient in a plane
is defined as 
\begin{equation}
A=\frac{1}{N}\sum_j e^{2i\theta_j},
\end{equation}
with $\theta_j$ the azimuth of particle $j$ in this
plane. Together with the
$2K_{\perp}/K_{\parallel}$ ratio it is widely used as a 
measurement of the
instability of a system. From this figure it is clear that all 3
methods indicate a similar result. For the models 
$r_a=1.0$ (right column) the
axis ratios and the $2K_{\perp}/K_{\parallel}(t)$
ratio remain nearly constant during the entire run, while the Fourier
coefficient either remains constant or increases upto a very low value
and then slowly declines, indicating no further evolution in the
stability has to be expected. Only the systems with $\mu=0$ and
$\mu=0.01$ have moderately increasing Fourier coefficients, indicating
a marginal, slowly evolving instability; models with higher $\mu$ can be 
considered stable.

Lower values of the anisotropy radius $r_a$ leads to unstable systems;
in columns 1 and 2 we display all systems with respectively $r_a=0.25$
and $r_a=0.5$. Clearly, the rate at which the bar is created depends
strongly on $r_a$. During a simulation of an unstable system the
$2K_{\perp}/K_{\parallel}(t)$ ratio rapidly reaches a maximum and then
declines to a value of $\approx 2.0$ after which it remains constant
for the remainder of the run. This maximum is an artifact of the
softening applied in the N-body simulation and is caused by the fact
that the initial conditions are not exactly in dynamical equilibrium
(see Sect.~2.3). Lowering the time-step and softening length caused a
diminution of the maxima. The time at which the
$2K_{\perp}/K_{\parallel}$ ratio reaches this constant value
corresponds with the time the axis ratios reach their lowest value and
the Fourier coefficients reach their highest value.  The influence of
the mass of an SBH is more clearly visible in the axis ratios and
Fourier coefficients. A remarkable result is the difference between
the systems $r_a=0.25$ and $r_a=0.5$ in the evolution of the axis
ratios: the former models first become triaxial, after which $b/a$
increases and $c/b$ decreases.  The latter models are prolate
axisymmetric during the entire run.

In summary, an SBH mass of a few percent can prevent or
reduce the bar instabilities in anisotropic systems. This result
agrees well with similar studies in disk galaxies \citep{norman, shen,
athanassoula, hozumi}. The effect is strongest for models
with strong radial anisotropies, where the decrease of the bar
strength is proportional to the SBH mass. In other words, while more
radially anisotropic systems develop stronger bars than more isotropic
models, the bars of the former are more easily affected by a
supermassive black hole (see Fig.~\ref{krkt}). This is to be expected,
since radial systems host more eccentric orbits, therefore more
particles from the outer regions pass near the center where their
orbits can be altered by the Kepler force of the SBH.\\

An alternative approach to present the evolution of the models is
given  
in Figs.~\ref{betasrho}-\ref{betasrho4}, where we show the evolution of 4
systems by means of the density $\rho(\bmath{r})$ and velocity
dispersions $\sigma_{\!\perp}(\bmath{r})$ and
$\sigma_{\parallel}\,(\bmath{r})$, in at dynamical times $t=0,\ t=25T_h$ and
$t=50T_h$. In each principal plane the moments are calculated on a
grid of 2500 points, with 50 nearest neighbors around every grid
position.

Fig.~\ref{betasrho} displays an Osipkov-Merritt system without an SBH
and anisotropy radius $r_a=0.25$. As was shown in 
Fig.~\ref{krkt} this model has a strong bar
formation, resulting into a new equilibrium state after $t=5T_h$
which it retains during the rest of the run (as can be seen at
$t=25T_h$ and $t=50T_h$).  This bar alters the density
distribution into a roughly triaxial symmetry, even peanut-shaped in
the $XZ$-plane where the radial instability is the most prominent. The
tangential dispersion $\sigma_{\parallel}$ increases
significantly. This occurs especially at the edges, where in contrast
the radial dispersion vanishes. This can be explained by the mechanism
of the bar formation: particles that pass through the bar are pulled
towards it, and eventually align their orbit with the bar.  Only the
orbits along the principal axes remain largely unaffected by the bar
due to the symmetric forces on these particles, hence their motion
remains radial.

In Fig.~\ref{betasrho2} a model with $r_a=0.25$ and $\mu=0.05$ is
shown.  Again a bar is formed, but less pronounced than in the absence
of an SBH.  Clearly, during the run the bar is reduced by the SBH,
causing a gradual increase in the $c/a$ axis ratio ($XZ$-plane). More
striking however is the evolution in the $XY$-plane, where the
ellipticity has disappeared. Thus, the model has become an oblate
axisymmetric system. This is also reflected in the dispersions:
$\sigma_{\parallel}$ again follows the bar structure, but the
cross-form $\sigma_{\!\perp}$ vanishes as particles pass near the
SBH. Since most particles reside in the $XY$-plane, on eccentric
orbits (since $r_a$ is small), the scattering in this plane is
strongest. After $t=50T_h$, we expect a further small increase in the
$c/a$ axis ratio, but as the velocity dispersion becomes more
isotropic fewer particles from the outer regions will pass near the center
(i.e.\ be affected by the SBH), hence the model will not change much further.

This can also be seen by comparing the system with $\mu=0.05$ to a 
model with $\mu=0.1$ (Fig.~\ref{betasrho3}). This model has essentially the 
same properties as the former. The larger SBH mass has above all influence on its
efficiency, resulting in a faster bar reduction.
 
Finally, we consider the effect of the anisotropy radius by analyzing
a system with $\mu=0.05$ and $r_a=0.5$
(Fig.~\ref{betasrho4}). Compared to Fig.~\ref{betasrho2}, the initial
bar is less strong, as expected. However, its structure and evolution
is different from the system with $r_a=0.25$.  First,
$\sigma_{\!\perp}$ remains spherically distributed during the run,
thus less scattering occurs. This implies less reduction of the bar
instability. Moreover, the density does not become symmetric around
the $Z$-axis. In contrast, the $X$-axis is now the symmetry axis
during the entire run, resulting in a prolate axisymmetric system. It
thus seems that models with an SBH become oblate or prolate, depending
on their velocity anisotropy. It would indeed be very interesting to
compare the orbital structure of both these systems in full detail.

As a final remark we note that inside a radius
$r_K=\sqrt{\mu}/(1-\sqrt{\mu})$ the force of the SBH is stronger than
the stellar component, so that all models remain spherical inside this
radius. In conclusion, systems with an SBH become axisymmetric systems
with a spherically symmetric core.

\section{Conclusions and summary}

Most mass estimates of SBHs result from dynamical models of either
stellar or gas kinematics. The inclusion of strong radial anisotropy
is considered in these models \citep{binney}, yet they have never been
tested for radial stability. Our goal was to test the stability of
systems with a central SBH and to look for any trend as a function of
the mass of the SBH. We used the same method that was previously
introduced by Meza \& Zamorano (1997) and extended it to systems with
a central SBH. We first tested the procedure on Hernquist systems
\citep{baes} without an SBH and with different anisotropic
behavior. Our method appeared to be efficient in discriminating the
stable from the unstable systems.

Instead of focusing on complicated numerically derived dynamical
models, we opted for analytical distribution functions that take the
effect of a central SBH into account, in order to be able to look for
any trend. Since the isotropic Hernquist models with an SBH do not
have distribution functions that are monotonically increasing
functions of the binding energy \citep{ciotti,baes2} and hence the
sufficient criteria of Antonov (1962) and Dor\'emus and Feix (1973)
for isotropic systems cannot be applied, we first investigated the
radial stability of these systems. No effect was found by letting the
mass of the SBH vary, giving only stable systems. However, in the case
of the anisotropic systems with an SBH we did find a dependence of the
stability of the system on the mass of the SBH. The more massive the
SBH, the more stable a system becomes, but especially the more the
instability is reduced. A trend which is most obvious in very
anisotropic systems (thus with very small anisotropy radius $r_a$). An
SBH with a mass of a few percent of the entire galaxy mass, is able to
weaken the strength of the bar, which is in correspondence with
similar studies in disk galaxies \citep{norman, shen, athanassoula,
hozumi}. Judging from Fig.~\ref{contourc}, the stability boundary of
$c/a \gtrsim 0.95$ over 50 dynamical times, shifts from $r_a \approx
1.1$ for $\mu=0$ to $r_a \approx 1.0$ for $\mu=0.1$. This corresponds
to $2 K_r/K_t = 2.2$ for $\mu=0$ and to $2 K_r/K_t = 2.0$ for
$\mu=0.1$. These values are in very good agreement with previous
authors.

Remarkably, systems with an SBH but with different anisotropy radii
$r_a$ evolve differently: highly radial systems first become triaxial
whereafter the SBH makes them more oblate if $\mu$ is large, while
less radial models tend to form first into axisymmetric prolate structures, 
that then become less elongated due to the influence of the SBH.

It is also interesting to note that the central density distribution
of systems with an SBH remains spherically symmetric during the entire
simulation out to a radius of half the effective radius. This is not
the case for systems without an SBH, which become axisymmetric or
triaxial, depending on $r_a$. Interestingly, this includes the region
that is considered for the $M_{BH}-\sigma$ relation, which predicts
such an evolutionary link between the central SBH and the spheroid
where it resides. Similarly, the central anisotropy parameter
decreases as a function of time at a rate proportional to the mass of
the SBH, due to more tangential orbits at the center.

Apart from a central SBH, one can also investigate the effect of
central density cusps \citep{sellwood01,holley01} or isotropic cores
\citep{trenti} on the radial stability. Pre\-vious research shows that both
central density cusps and isotropic cores act as dynamical
stabilizers, hence the same effect as the addition of an SBH. In the
future we plan to look at the combination of both investigations,
namely the combination of an SBH and central density cusp.

Ultimately, one can of course verify the radial stability of the
state-of-the-art models that are being used to estimate the mass of
the SBHs with e.g.our methodology, however this investigation was not
the goal of this paper.

\section*{Acknowledgements}
The authors would like to thank L. Ferrarese and the people of the
Herzberg Institute of Astrophysics in Victoria for the very fruitful
discussions. SDR and PB are postdoctoral fellows with the National
Science Fund (FWO-Vlaanderen).

\appendix
\section{Hypergeometric function}
A relevant definition of Generalized hypergeometric functions
$_pF_q(a_1,\ldots a_p;b_1,\ldots,b_q;x)$ can be found in Gradshteyn \&
Ryzhik Sec.\ 9.14, page 1071 in the 5th edition. For specific arguments
hypergeometric functions can be reduced to more simpler analytical
functions, this can be done with mathematical software packages as
e.g.\ Maple or Mathematica. For general coefficients however, the
evaluation needs to be done by means of hypergeometric series:
\begin{eqnarray}
_pF_q(a_1,\ldots,a_p;b_1,\ldots,b_q;x)&=&1+\sum_{j=0}^{+\infty}\\\nonumber
&&\prod_{k=0}^j\frac{x\prod_{i=1}^p(a_i+k)}{(1+k)\prod_{i=1}^q(b_i+k)}.
\end{eqnarray}
This routine is very suitable to numerically evaluate 
any generalized hypergeometric function within a certain degree of
accuracy.

\label{lastpage}

\end{document}